\documentstyle{article}
\begin{document}
\newcommand{\be}{\begin{equation}}
\newcommand{\ee}{\end{equation}}
\newcommand{\bc}{\begin{center}}
\newcommand{\ec}{\end{center}}
\newcommand{\fr}{\frac}

\centerline{\huge\bf A BRIEF HISTORY OF THE EARTH}
\vspace{2in}
\centerline{\large \bf ARBAB I.ARBAB\footnote{E-mail:
arbab@ictp.trieste.it,
arbab64@hotmail.com, arbab@pan.uzulu.ac.za}}
\vspace{1.5cm}
\centerline{The Abus Salam International Center for Theoretical Physics,
P.O. Box 586, Trieste 34100, ITALY}
\vspace{2cm}

\centerline{ABSTRACT}
\vspace{1cm}
{\large The possibility of an increasing gravitational constant ($G$) and its
implications on the earth's history is discussed. The model is consistent with geophysical and astronomical
data. The number of days in a year in early epochs predicted by  the model is in agreement with those
obtained from radiometric and fossil corals. The model considers the earth to be cold
in the beginning and it gradually heats as time evolves. Paleontology, paleotemperature,
paleogeophysics and paleobotany data are needed to restrict the law of variation
of $G$. The model predicts that the earth has undergone an evolutionary process
resulting in its present phase.

\vspace{2cm}

Heading: Geophysics; Astronomy; Pale-ontology; Earth Science
\newpage
\large
{\bf INTRODUCTION}\\

In 1937 Dirac [1] proposed a model in which the gravitational constant $G$ decreases
with time as $G\propto 1/t$, but his model failed when confronted with observations.
The possibility of an increasing $G$ has recently been considered by several workers [2,3,4].
In both cases the variation of $G$ is of the order of the Hubble constant ($H$).
According to the theory of general relativity (GR) the gravitational constant is constant.
Hence any theory requires a change of $G$ would imply a change in GR.
If $G$ turns out to be varying with time then GR should be revised. The supporters
of GR however are reluctant to accept the idea of a varying $G$.
\\
We will in this paper investigate the geologic implications  of an
increasing gravitational constant $G$ on the earth.\\
The best experimental upper limit on the present rate of change
of the gravitational constant $G$ comes from the analysis of radar observations of Mercury and Venus.
For a planet in a circular orbit, around a central mass ($M$), of radius $r$ and velocity $v$, we have $MG=v^2r$,
 so if the orbital angular momentum $mvr$ stays fixed while $G$ changes, then $r$ and $v$ will
 vary as [5]
 \be
 r\propto 1/v\propto 1/G
 \ee
 and the orbital period $T=2\pi r/v$ will vary as
 \be
 T\propto 1/G^2
 \ee
 The motion of a body moving under gravitational attraction would be affected
 by the presence of a resisting medium. For a circular motion, this was shown
 to result in an increase in the orbital velocity and a decrease in the orbit ($r$) [6].
 By repeated comparison of the orbital periods of the inner planets over the period
  1966-69, with time as told by an atomic clock (which does not depend on $G$), Shapiro
   {\it et al} have set an upper limit
   \be
   |\dot G/G|<4\times 10^{-10}\ \rm year^{-1}
   \ee
   This is almost good enough to rule out Dirac theory, but is not yet sufficiently
     stringent to put a useful limit on Brans-Dicke [7] theory coupling parameter
$\omega$.
   There was also a prospect of setting an upper limit on the rate of change of $G$
   from analysis of the flight time of laser signals, sent from the earth to the moon,
   and reflected back to earth by the corner reflectors placed on the moon's surface
   by the Apollo expedition.
   Variation in $G$ over the last five millennia can perhaps be determined from
   the study of ancient eclipse records.
   It may be possible to measure changes over the last 350 million years
   in the number of days in  lunar month or a year, by counting monthly or
   annual growth bands and daily growth ridges on fossil coral.
   With decreasing $G$, the radius of the earth would increase roughly as $G^{-0.1}$
   , causing complicated damage to the earth's crust.
   If $G$ were large in the past, then  stars would have run their thermonuclear evolution
   more rapidly. If $G$ were greater in the past, then the sun luminosity $L$ would
   have been greater by a factor roughly proportional to $G^8$, and that the radius
   $r$ of the earth orbit would have been smaller by a factor proportional to $G^{-1}$, so that
   the surface temperature of the earth, which varies  more or less as
   $(L/r^2)^{1/4}$, would have been greater by a factor $G^{2.5}$.
   If $G$ decreases as $t^{-0.09}$, (Brans-Dicke, $\omega=6, \ G\propto t^{-2/(4+3\omega)}$) and if the age of the universe
   is $8\times 10^9$years, then the temperature of the earth's surface $2\times 10^9$ years ago would
   have been only $20^o\rm C$ higher than at present, which need not have had any drastic
   effect on biological evolution. On the other hand, if $G$ has decreased as $1/t$
   (Dirac), then the temperature of the earth's surface $10^9$ years ago
   would have been above the boiling point of water unless the earth's albedo
   was very much higher than at present.
   Thus too large an early value of $G$ could have  prevented the evolution of life
   forms capable of invoking curiosity about the universe [8,5].\\
   If $G$ has varied with time, the geologic consequences will have been fundamental.
   For many years the grounds for supposing that $G$ changes with time were entirely
   philosophical, but there is now some physical evidence that this conclusion
   is correct.\\

{\bf The Model}\\

   Recently, a cosmological model based on Einstein theory  has been proposed [9,10].
   The model considers that the gravitational constant is increasing with time as
   $G\propto t^{(2n-1)/(1-n)}$, where $n$ is the viscosity index ($\frac{1}{2}<n<1$)
   and that the age of the present universe is $t_p=\frac{1}{3(1-n)}H^{-1}_p$, where
   $H_p$ is the Hubble constant.\\
   For $n=0.7$ , $G\propto t^{1.3}$ and $t_p=1.1 H^{-1}_p$. This gives $\dot G/G=1.2H^{-1}_p$.
  The radius of the earth ($R$), temperature $(\theta$), orbital velocity ($v$)
   , orbit ($r$), orbital period ($T$), escape velocity ($v_e$),
   and the acceleration due to gravity ($g$) vary as [3,4]:\\
\begin{equation}
 R\sim G^{-0.1}, \ \theta\sim G^{2.5}, \ T\sim G^{-2},  \ v\sim G, \ r\sim G^{-1},\
   v_e\sim G^{0.55}, \ \ g\sim G^{1.2}\ .
   \end{equation}
   Based on these relations one can relate these  parameters at two different
   epochs viz.\\
\begin{equation}
\frac{R_a}{R_p}=(\frac{t_p}{t'})^{0.13}, \frac{\theta_a}{\theta_p}=(\frac
   {t'}{t_p})^{3.25},\ \frac{T_a}{T_p}=(\frac{t_p}{t'})^{2.6}, \
\frac{r_a}{r_p}=(\frac{t_p}{t'})^{1.3},\  \frac{v_a}{v_p}=(\frac{t'}{t_p})^{1.3}, \ \ \ \frac{g_a}{g_p}=(\frac{t'}{t_p})^{1.56}
 \ ,\frac{v_{e\ a}}{v_{e\ p}}=(\frac{t'}{t_p})^{0.715}
   \end{equation}
   where $t'=t_p-t_a$, is the time measured from the beginning of the universe and
   the subscript `a' and `p' refer to the value of the parameter before the
   present  and at the present epochs, respectively.\\
   Of primary importance, the model predicts that the temperature of the earth was
$266 \rm K$ 400 millions ago while Dirac model gives $424\rm K$. Hence our model predicts
that life could have been possible in that epoch.
The temperature-time relationship is plotted in Fig.(1), it shows the slow warming of the earth.
It was suggested by [11,12] that the earth probably formed from accretion of cold
planetetismals  and therefore was initially homogeneous. However, shortly after formation
, the earth must have heated up so that it could differentiate or separate into a core, mantle, and crust.
Heating was the result of heat generated by impacts during accretion, by gravitational
 compression, and by radioactive decay.
 Each in-falling planetesimal carried a great deal of kinetic energy due its motion,
a large part of which was converted to heat upon impact.
 The rate of heat generated in this manner depended upon the rate of accumulation
 and the velocity of planetesimal. As material accumulated on the surface of the growing planet
 the inner parts of the planet were subjected to higher and higher pressure which resulted
 in a proportional increase in temperature. Because rocks are poor conductors of
 heat the temperature of the earth's interior continued to rise. At the same point iron
began to melt and because of its high density began to sink towards the center of the planet,
forming the core. In Kuiper's model [13] the earth is formed from accumulation
of materials in a cold ($<-200^o\rm C$) nebula. However, our model gives a value of $-218^o\rm C$.
At that time the earth was at roughly double its distance from the sun now. The
earth was in the position occupied by Mars some 3050 million years ago,
therefore Mars today could represent  an earlier life's  history  of the
earth. Therefore, the knowledge of Mars life conditions at the present time will
help speculate about the history of the earth. \\ 
Because the variation of $G$ is related to the age of the universe and to the deceleration
parameter of the universe, the geologic constraints for $G$ would improve the
knowledge about the present universe.
Equation (5) can be inverted to give a geologic time scale of the earth (calendar).

Inserting a present age of the universe ($11\times 10^{9}\rm y$) we, from eq.(5),
tabulate the earth data (Table 1) for the different geologic periods.\\
Wells [14], using coral as an illustration, presented the idea that the fine
growth lines represented daily increments.
He was able to infer the number of days in a year the time the coral lived.\\
He noted that Astronomers believed the period of rotation of the earth around its
polar axis has been gradually
slowing down due to a slight damping effect of tidal forces on rotational energy,
and that in the course of  geologic time the length of the day has been
increasing and the number of days in the year decreasing [15].\\
In our present model the length of the day (D) is varying as $D\propto G^2$ or
$D\propto t^{2.6}$.
It is therefore clear that the length of the day was 21.9 hours 380 million years
ago [12,14]. This lengthening of the day is possibly due to the tidal effects of the moon.
Algal Stromatolites, similar to types
that today form chiefly in the inter-tidal zone were several centimeters high,
suggesting that the moon was present at least by then to influence tides. The increase
in heights of these Stromatolites would suggest a gradual decrease in the distance
of the moon from the earth as the cause of increased tides [17].

Fossils have great intrinsic interest as evidence of changes that the life of
the earth has undergone of the evolutionary process that has now produced the
present biological environment. It was well known that the moon has influence
on biological processes and that many biological cycles are related to the
phases of the moon [12].\\
Some experts attribute the rather high "mortality" in marine invertebrates in Permian
time  to cold climates and shrinking seas. By the same token the extinction of dinosaurs
is likely to be due to environmental changes and not due to any global catastrophe
as many people suggested.\\
Astronomical data indicate that the Cambrian year should have about 420 days,
so that 400 days for the Devonian year in line with the astronomical calendar based
on the theory of decelerating rotation.
It was found that ages inferred from  fossil coral ( biological system)
generally agree with the radiometric time.\\
Wells emphasized that if more detailed studies supported his ideas, fossils with patterns of additive growth
were potentially useful chronometers of absolute time [14,16].\\
Plots of the temperature, number of days per year, earth's radius, earth's orbit
and the length of day {\it versus} time  are shown
in  Fig.(1), Fig.(2), Fig.(3), Fig.(4) and Fig.(5), respectively.
\\
  Working experimentally with the living quahog {\it Mercenaria}, Panella and Mac Clintock established that the
basic growth units of the shell were, indeed, daily increments. Moreover, they found a number
of distinct cycles patterns of increments, the most obvious being cycles reflecting
seasonal changes in series of thicker (summer) and thinner (winter) increments.
The $\rm O^{18}/\rm O^{16}$ ratios determined for successive concentric zones in that
shell indicate rhythmic fluctuations of temperature from a winter low of about
$15^{o}\rm C$ to a summer high of about $22^{o}\rm C$. From these data it was deduced
that animal lived some 140 million years ago [16].
Comparing this with our prediction we get a temperature of $14.7^o\rm C$.

The number of days per year in the past (Table 1), according to the present model,
shows a complete agreement with the data obtained from fossil coral and
radiometric time (Table 2).\\
According to our model, much of the internal heating of the earth is presumed to
have occurred subsequent to its aggregation through the gravitational attraction and
radioactive decay within the mass. The energy produced by such heating apparently
has caused the continual disturbances seen in earthquakes, volcanos, formation of
mountain, and the like.

 During the 19th century, geologist sought an explanation for diastrophism in a shrinking earth.
 It was then widely believed that the internal heat is a residue from a primordial molten
 stages and that as it slowly cooled the earth began to solidify at the surface, thus
 developed a rigid crust [15].
 We, again, present the idea of a shrinking earth but on  different ground. In our model
 the solidification of the crust is due to an
increase in the strength of gravity with time. One of the immediate consequences of
the shrinking earth is that the sea-covering areas would increase and the continent would
shrink as time evolves.
Using the Hubble constant $H_p=10^{-10}\rm y^{-1}$, the age of the universe
$t_p=11\times10^9\rm y$, the deceleration parameter $(q_p$) of the universe
is  $q_p=0.1$.\\

For a planet to retain its atmosphere over billion years the escape velocity
must be such that $v_{rms}=10v_e$, where $v_{rms}$ is the root mean square velocity of a given species.
Using the above relation, the atmospheric temperature $T_{atm}$ is related to $G$ by
$T_{atm}\sim G^{1.1}$. \\
For instance, water under cool conditions react to form
compounds that do not easily evaporate, where as neon is chemically inert and
remains
a gas. Thus a cool primitive earth would selectively retain water (and other
chemically reactive gas like carbon dioxide and ammonia) by locking them up as
compounds in rock, where as inert gases like neon would gradually escape. Thus
life could have developed some billion years ago.

We remark that the appropriate law of variation of $G$ has to be found from
the variation of the geologic parameters (earth's radius, temperature, speed, etc.)
with time. We have, in this model, highlighted the implications of an increasing gravity on the
evolution of the earth.
\newpage
{\bf Concluding Remarks}\\

Thus, as a result of increasing gravity, the earth continues to shrink and speeds up
in its orbit while approaching the sun. Like the astronomical calendar which is based on
the theory of decelerating rotation, our model based on an increasing gravity
can be used as a calendar for the geologic time. The model predicts events that
took place in the initial state of the earth formation. We also predict that the
sea-covering areas are increasing while the continental areas are decreasing.
Using the present age (Hubble) parameter as an input one can predict the
value of the Hubble (age) parameter of the universe.
Paleotemperature, paleoclimatology, paleogeophysics, paleogeography data
will set appropriate limits on the variability of $G$. The consequences
of this variation will be very substantial in understanding the history of
the earth.
\\

{\bf References}\\
\\
1- P.A.M. Dirac, {\it Nature 139},323, 1937\\
2- L.S.Levitt, {\it Lett. Nuono Cimento. 29}, 23, 1980\\
3- A.-M.M.Abdel Rahman, {\it Gen. Rel. Gravit. 22}, 655, 1990\\
4-C.Massa, {\it Astr. Phys. Space. Sci. 232}, 143, 1995\\
5- S.Weinberg, {\it Gravitation and Cosmology}, pg. 619, pg.629 , John Wiley and Sons Inc., 1972\\
6- Harold Spencer Jones, {\it General Astronomy}, pg.226, Edward Arnold Publishers Ltd, 1961\\
7- C.Brans and R.H.Dicke, {\it Phy. Rev. D124}, 203, 1961\\
8- M.V. Berry, {\it Principles of Cosmology and Gravitation}, pg.136, IOP Publishing Ltd 1989\\
9- A.I.Arbab, {\it Gen. Rel. Gravit., 29},61, 1997\\
10- A.I.Arbab, {\it Nouvo. Cimento B113}, 403, 1998\\
11- D.C.Tozer, {\it Science Progress 64}, 1, 1977\\
12- Billy P.Glass, {\it Introduction to Planetary Geology}, pg.43, pg.87, Cambridge University Press, 1982\\
13- G.P. Kuiper, Astrophysics, ed. J.A.Hynek, pg.404, New York McGraw-Hill (1951)\\
14- J.W. Wells, {\it Nature V.197, Nr.4871}, 948 , 1963\\
15- Don L.Eicher, {\it Geological time}, pg.138, Prentice - Hall, 1968\\
16- Carl O.Dunbar, {\it Historical Geology}, pg.67, pg.75 John Wiley \& Sons Inc., 1969\\
17- Jr. Robert H.Dott and R.L.Batten, {\it Evolution of the Earth}, pg.288-289
MacGraw-Hill Company, 1981
\newpage
\begin{table}
\caption{Data obtained from the {\it principle of increasing gravity}}
\vspace{1cm}
\begin{tabular}{|r|r|r|r|r|r|r|r|r|r|r|r|}
\hline
time* & 65 & 190 & 280 &  345 & 570 & 1000 & 2000 & 3000 & 4500\\
\hline
length of day(hr) & 23.63 & 22.94 & 22.44 & 22.10 & 20.90 & 18.73 & 14.24 & 10.49& 6.11\\
\hline
period** & 370.7 & 381.9 & 390.3 & 396.5 &419.2 & 467.6 &615.0 & 838.3 & 1433.3 \\
\hline
temperature (K)& 294.3 & 283.5 & 275.9 & 270.5 & 252.2 & 220.1 &165.2 & 106 & 54.2\\
\hline
radius(km) & 6382.9 & 6392.5 &  6399.4 & 6404.5 &6422.3 & 6457.5 &6546.5 & 6647.5 & 6829.2\\
\hline
escape velocity (km$s^{-1})$& 11.15 &11.06 & 11.00 & 10.95 & 10.78 & 10.46 & 9.70 & 8.92 & 7.69\\
\hline
orbit(x$10^{11}\rm m$)&1.51 & 1.55 & 1.58 & 1.61 & 1.69 & 1.69 & 1.94 & 2.26 & 2.96\\
\hline
\end{tabular}
\end{table}
\begin{table}
\caption{Data obtained  from {\it fossil corals and radiometric time}}
\vspace{1cm}
\begin{tabular}{|r||r|r|r|r|r|r|r|r|r|r|r|}
\hline
time* & 65 & 136 & 190 & 225 & 280 & 325 & 345 & 395 & 435 & 500 & 570\\
\hline
period** & 371 & 377 & 382 & 385 & 390 & 394 & 396 & 401 & 405 & 412 & 421\\
\hline
\end{tabular}
\end{table}
[*: Million years before present, **: Number of days per year]

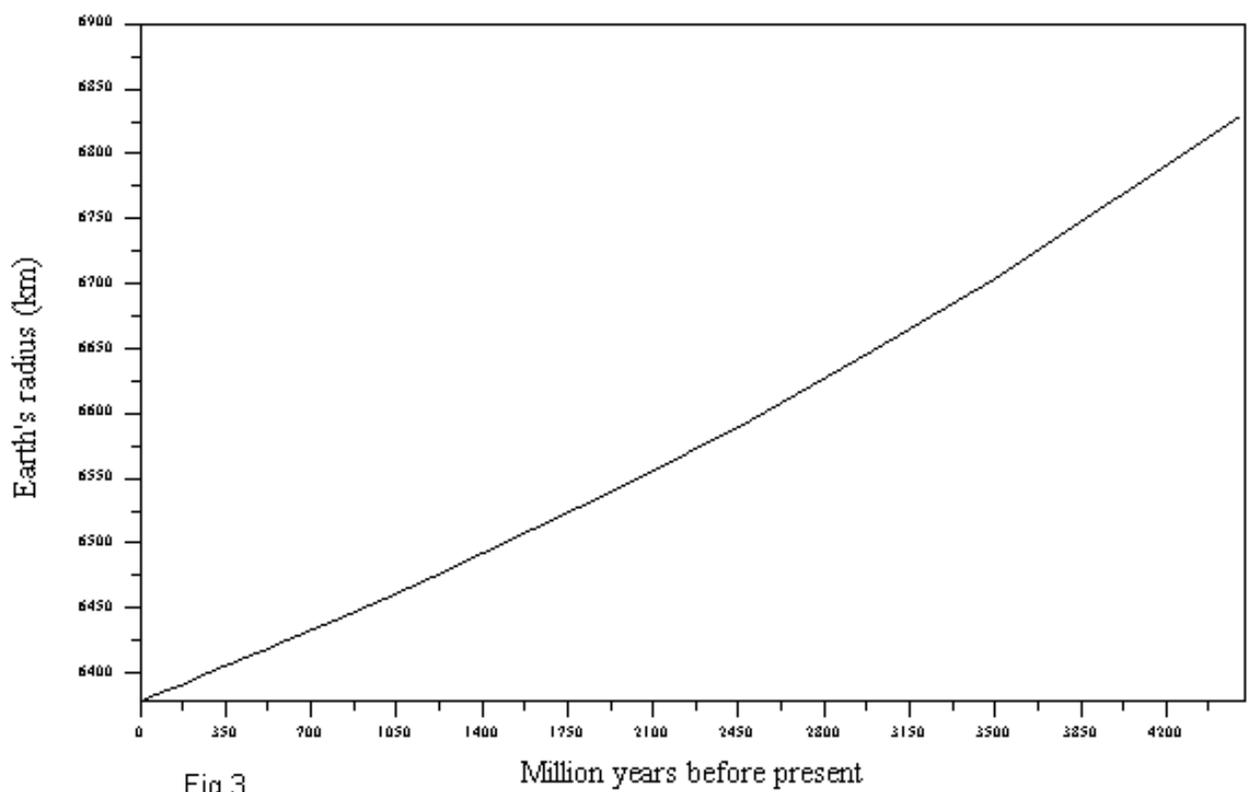
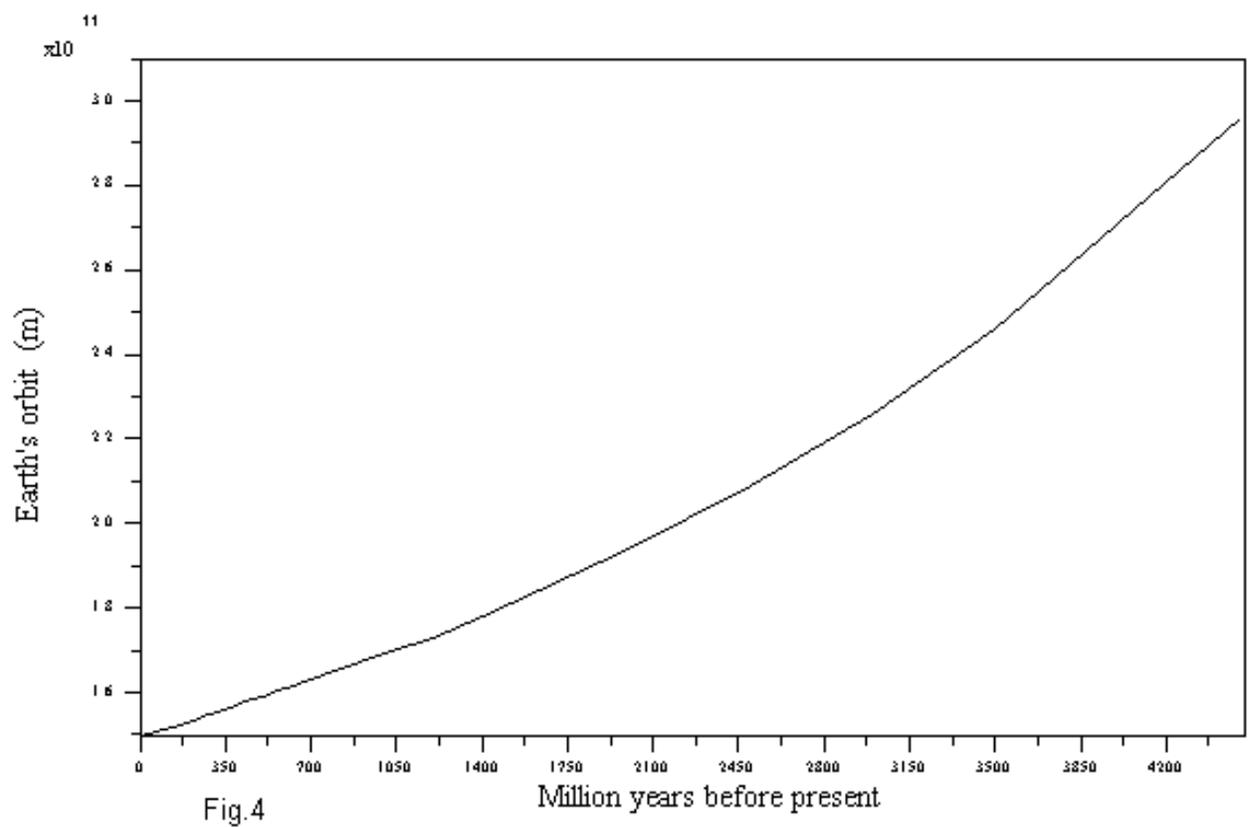
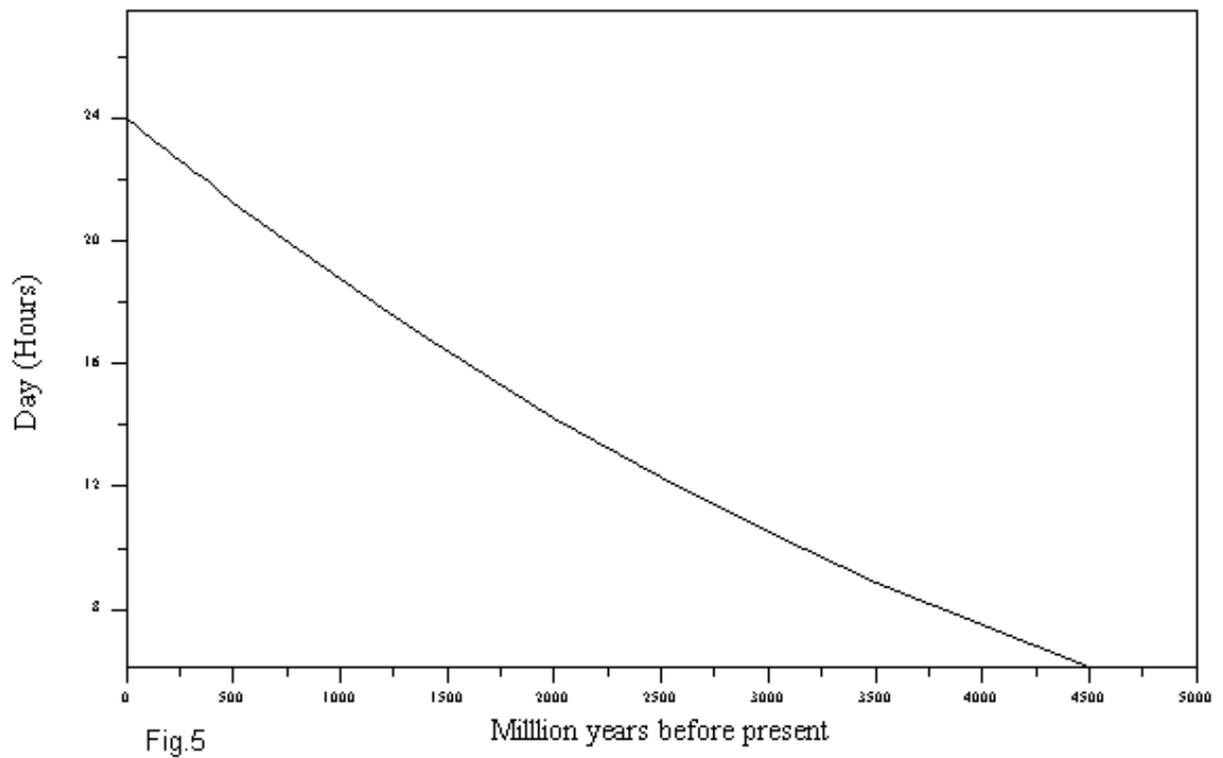
\end{document}